\documentclass[aip, pop,  reprint]{revtex4-1}

\usepackage{amssymb}
\usepackage{amsmath}
\usepackage{graphicx}
\usepackage[T2A]{fontenc}
\usepackage[utf8]{inputenc}
\usepackage[english]{babel}

\renewcommand{\vec}[1]{\mathbf{#1}}
\newcommand{\rem}[1]{}
\newcommand{\HH}{\mathcal{H}}
\newcommand{\HP}{\mathcal{P}}

\newcommand{\rmi}{\mathrm{i}}
\newcommand{\rme}{\mathrm{e}}
\newcommand{\p}{\partial}
\newcommand{\ee}{\varepsilon}
\newcommand{\x}{\vec{x}}
\newcommand{\q}{\vec{q}}
\def\sign{\mathop{{\rm sign}}}
\def\Real{\mathop{{\cal R}{\rm e}}\nolimits}
\def\Imag{\mathop{{\cal I}{\rm m}}\nolimits}

\begin{document}

\title{Quasi-optical theory of microwave plasma heating in open magnetic trap}
\author{A. G. Shalashov}
\email[Author to whom correspondence should be addressed. Electronic mail:\;]{ags@appl.sci-nnov.ru}
\author{A. A. Balakin}
\author{E. D. Gospodchikov}
\author{T. A. Khusainov}
\affiliation{Institute of Applied Physics of the Russian Academy of Sciences, Ulyanova str. 46, 603950 Nizhny Novgorod, Russia}
\affiliation{Budker Institute of Nuclear Physics, Siberian Branch of the Russian Academy of Sciences, Akademika Lavrentieva ave. 11, 630090 Novosibirsk, Russia}

\date{\today}
\begin{abstract}
Microwave heating of a high-temperature plasma confined in a large-scale open magnetic trap, including all important wave effects like   diffraction,  absorption, dispersion and wave beam aberrations, is described  for the first time  within the  first-principle technique  based on consistent Maxwell's equations. With this purpose,  the quasi-optical approach is generalized over weakly inhomogeneous gyrotrotropic media with resonant absorption and spatial dispersion, and a new form of the integral quasi-optical equation is proposed. An effective numerical technique for this equation's solution is developed and realized in a new code \textsl{QOOT}, which is verified with the simulations of realistic electron cyclotron heating scenarios at the Gas Dynamic Trap at the Budker Institute of Nuclear Physics  (Novosibirsk, Russia).
\end{abstract}
%\pacs{42.25.Fx, 02.70.Bf, 02.60.Cb}

\maketitle

\section{Introduction}

The absorption of electromagnetic waves under the electron cyclotron resonance (ECR) conditions is widely used for heating of high-temperature plasmas in toroidal magnetic traps (tokamaks and stellarators). However, for many years the use of this method in open magnetic configurations has been limited either by plasma heating in relatively compact laboratory installations\cite{V16}, or by MHD stabilization of low-density plasma\cite{V13,V14,V15}. The only exception was the TMX-U experiment at Lawrence Livermore National Laboratory where electron temperatures up to 0.28 keV were obtained with ECR heating, but these studies were concluded soon\cite{V12}. And only recently an efficient ECR heating of a dense   (comparable to toroidal devices) plasma has been demonstrated at a large-scale mirror trap. We mean successful experiments on a combined plasma heating by neutral beams and microwave radiation performed at the Gas Dynamic Trap  (GDT) device in the Budker Institute of Nuclear Physics\cite{X7,X8,X9,X10}.  In particular, direct and highly-localized heating of the thermal electron component allows to achieve  the GDT experiment the record for open traps electron temperature of 1 keV  at the plasma axis\cite{X8}. These studies convincingly demonstrate good prospects for the use of simple axially symmetric open magnetic traps as a powerful source neutron  for fusion applications \cite{Simonen}.

Implementing  the effective ECR heating of a dense plasma at a large open trap requires a revision of the prevailing ideas about the physics of cyclotron absorption as well as the subsequent transport   of energy and MHD stabilization of a plasma column. None of the well-understood techniques developed for the toroidal plasma heating works well in this case \cite{X4}. Numerical modeling of the propagation and absorption of electromagnetic waves in an inhomogeneous plasma plays an important role in this occasion. Until recently, such modeling was only possible in the framework of the geometric optics approximation also known as ray-tracing. In particular, the basic ECR heating scenario used in the GDT has been originally proposed and justified with the ray-tracing calculations \cite{X4,X5,X6}. However, in this scenario there are areas of reflection and strong absorption of waves, in which the medium is not smooth as compared to a wavelength. So, detailed understanding of the involved processes requires to go beyond the geometric optics approximation.

The main effects that violate the geometric optics are associated with the spatial dispersion in a strongly inhomogeneous region of resonant absorption, the diffraction of a wave beam, and the caustic formation in the vicinity of internal reflection points of a wave beam. Straightforward simulation of these effects for large devices within a complete set of Maxwell's equations is very complicated, in particular, because of the smallness of a wavelength. A good alternative is the consistent quasi-optical approach based on an asymptotic expansion of Maxwell's equations in the paraxial approximation in the vicinity of the selected Wentzel--Kramers--Brillouin (WKB) mode \cite{Balakin_2007a,fullQOp2}.

In this paper, the quasi-optical approximation is adopted to describe the propagation of wave beams in a high-temperature plasma in an open magnetic trap. A similar approach was previously developed for  toroidal magnetic traps \cite{Balakin_2008,BECCD}. However, description of the microwave heating in modern open traps needs substantial modification of the quasi-optical theory due to a requirement of more accurate accounting of the spatial dispersion inside the resonant absorption zone. The physics of this difference originates from the fact that the magnetic field in open configuration typically changes along its direction, while in the toroidal geometry the magnetic field varies presumably in orthogonal direction. As a result, the factorizing approach to the quasi-optical Hamiltonian, that works well in the modeling of wave propagation in a toroidal trap, fails completely in case of a large open trap.  This motivates us to find a new, more general and accurate, non-factorized formulation of the theory resulted in the substantial modification of a numerical algorithm, which eventually led to development of an entirely new code.

The remainder of this paper is organized as follows: Sec.~II derives  the basic equations of quasi-optical approximation and their applicability conditions;  Sec.~III  defines a general ansatz for a  dielectric response of inhomogeneous dispersive media and derive the quasi-optical evolution operator; peculiarities related to wave dissipation are considered, and modification of the dissipative part of the early defined evolution operator is proposed in Sec.~IV; a numerical method for solving the quasi-optical equation is developed in Sec. V; the  model is adopted to a particular case of hot magnetized plasmas confined in an open magnetic trap  in Sec.~VI; first results of simulations of real scenarios of ECR plasma heating at the GDT device are discussed in Sec.~VII; and Sec.~VIII  summarizes the results.

\section{Basic quasi-optical equation}

A rigorous derivation of the quasi-optical approximation for a weakly inhomogeneous gyrotropic and anisotropic media with spatial dispersion may be found in  Refs.~\onlinecite{fullQOp1,fullQOp2}. The main steps are described below. Let us start with Maxwell's equations for $\rme^{-\rmi \omega t}$ processes written in the ``operator'' form
\begin{equation}\label{eq_maxwell}
\hat{L}_{ij}[E_j(\vec{r})]=0,\quad\hat{L}_{ij}\equiv \delta_{ij}\hat{k}^2-\hat{k}_i\hat{k}_j-k_0^2\hat{\ee}_{ij},
\end{equation}
where a summation over double indexes is implied. The operator $\hat{L}_{ij} [E_j (\vec{r})]$ acts on $j$-th Cartesian component $E_j (\vec{r})$ of the electric field vector, $k_0 = {\omega}/{c}$ is the vacuum wave number, $\hat{\vec{k}}$ is the wave number operator defined as the differentiation in the coordinate space, and $\hat{\varepsilon}_{ij}$ is the linear operator obtained from the dielectric tensor $\varepsilon_{ij}(\vec{r}, \vec{k})$ by a formal substitution $\vec{k}\to \hat{\vec{k}}$,  
$$\hat{\vec{k}}=-\rmi \, \nabla,\quad \hat{\varepsilon}_{ij}={\varepsilon}_{ij}(\vec{r},\hat{ \vec{k}}).$$
The latter operation is not uniquely defined because $\vec{r}$ and $\hat{ \vec{k}}$ do not commute; in the next section we describe the ``best way'' of regularizing the expression for $\hat{\varepsilon}_{ij}$ providing that the kernel ${\varepsilon}_{ij}$ is known. The result may be expressed in terms of Fourier-integral:
\begin{multline}\label{eq:e}
\hat{\varepsilon}_{ij}[E_j(\vec{r})] = \int \frac{\varepsilon_{ij}(\vec{r},\vec{k}')+\varepsilon_{ij}(\vec{r}',\vec{k}')}{2}\;\times\\ \times \rme^{\rmi  \vec{k}' (\vec{r}-\vec{r}')} E_j(\vec{r}') \frac{d\vec{r}' d\vec{k}'}{(2\pi)^3}.
\end{multline}

For the smoothly inhomogeneous medium, the approximate solution of Maxwell's equations can naturally be  found in the form of quasi-optical beam corresponding to some chosen WKB mode. To do this, we define the polarization vector $\vec{e}$ for the interested WKB mode as a solution of algebraic Maxwell's equations  $L_{ij}e_j=0$ in a locally homogeneous medium. Next, using the same technique as in the transition ${\varepsilon}_{ij} \to \hat{\varepsilon}_{ij}$, the polarization vector $\vec{e}(\vec{r}, \vec{k})$ can be associated with the polarization operator $\hat{\vec{e}}(\vec{r}, \hat{\vec{k}})$. Then, the approximate solution of Eq.~\eqref{eq_maxwell} can be found in the following form
\begin{equation*}\label{eq_sol}
E_j(\vec{r})=\hat{e}_j(\vec{r},\hat{\vec{k}})[U(\vec{r})].
\end{equation*}
Here, the polarization operator acts on the scalar amplitude  $U(\vec{r})$ of the wave beam. The equation for $U$ takes the form
\begin{equation}\label{eq_amplitude}
\hat{\HH}(\vec{r},\hat{\vec{k}} )[U(\vec{r})]=0,\quad \hat{\HH}\equiv\hat{e}^*_i\hat{L}_{ij}\hat{e}_j.
\end{equation}
Note that the corresponding \emph{function } $$\HH_{GO}=\Real\HH(\vec{r},\vec{k})$$ is usually used as a ray Hamiltonian in the geometrical optics. Similar, we will call the \emph{operator} $\hat{\HH}$ as a quasi-optical Hamiltonian.

Presently, we consider  axisymmetric open traps and magnetic configurations with weakly broken axial symmetry. In this case, we can use either Cartesian coordinate system or a cylindrical coordinate system with the polar axis aligned along the trap axis. Let us introduce the coordinates and the canonically conjugated momenta as
\begin{equation}\label{eq:coords}
\vec{r} \equiv (\vec{x},z),  \quad \hat{\vec{k}} \equiv (\hat{\vec{q}},\hat{k}_z),
\end{equation}
where $z$ is the coordinate along the trap axis, $\vec{x}=\vec{r}_\perp$ is the set of two coordinates in the transverse plane, $\hat{\vec{q}}=-\rmi\nabla_\perp$ and $\hat{k}_z=-\rmi \p/\p z$.  

Assuming the smoothness of beam parameters along $z$, we can introduce the scalar amplitude of $U(\vec{r})$ in the form
\begin{equation*}\label{eq_amp}
U(\vec{r})=\tilde{u}(\vec{x},z)\exp\left(\rmi \int \kappa(z) \,dz\right),
\end{equation*}
where $\tilde{u}$ is the complex envelope of a wave beam, and  function $\kappa(z)$ defines the dependence of the ``carrier phase'' of the wave field along the axis. Substituting this field in Eq.~\eqref{eq_amplitude}, we obtain the equation for the  envelope $\tilde{u}$  as
%\cite{Balakin_2008, Balakin_2007a, fullQOp2}:
\begin{equation}\label{eq_qo1}
\hat{\HH}\left(\vec{r},\hat{\vec{k}}+\kappa(z)\,\vec{z}_0 \right)[\tilde{u}(\x,z)]=0.
\end{equation}
In contrast to  geometrical optics, here we have certain freedom in the choice of $\kappa(z)$ related to the phase variation over the transverse aperture of the wave beam. After several tries, we found that the most convenient and reliable way in numerical calculations is  to define $\kappa(z)$ as a solution to a local dispersion equation with transverse wave vector $\vec{\tilde{q}}(z)$ corresponding to the center of mass  of a wave spectrum in the cross-section $z$:
\begin{equation*}%\label{eq:locdisp}%notused
	 \HH_{GO}\left(\vec{\tilde r}, \vec{\tilde q}(z)+\kappa(z)\,\vec{z}_0\right) = 0,\quad \vec{\tilde r}=(\vec{\tilde x},z),
\end{equation*}
where $\vec{\tilde x}$ is the  center of mass  for a wave field in the cross-section $z$. 

Assuming the complex envelope $\tilde{u}$ to be a smooth function  on the longitudinal coordinate, $|\p \tilde{u}/\p z| \ll \kappa |\tilde u|$, we expand Eq.~\eqref{eq_qo1} to the first order in powers of $\hat{k}_z$ (or, equivalently, in powers of $\p \tilde{u}/\p z$):
\begin{eqnarray} \label{H2}
\hat{\HH}[\tilde{u}] \approx \hat{\HH}_0[\tilde{u}] -\rmi  \frac{\p \hat{\HH}_0}{\p \kappa}\left[\frac{\p \tilde{u}}{\p z}\right] - \frac{\rmi}{2} \frac{\p^2 \hat{\HH}_0}{\p \kappa \p z}[\tilde{u}]  = 0,
\end{eqnarray}
where $\hat{\HH}_0=\hat{\HH}(\vec{r},\hat{\vec{q}}+\kappa\,\vec{z}_0)$, i.e. it is  $\hat{\HH}$   with omitted  derivatives over $z$.
With the limit of geometrical optics, the last term in  Eq.~\eqref{H2}  results in a pre-exponential factor responsible for the variation of the wave amplitude due the effect of  ``group'' slowing down. This term may be eliminated with formal substitution
$$\tilde{u} = \hat A[u], \qquad \frac{\p \hat{\HH}_0}{\p \kappa}\frac{\p \hat{A}}{\p z} + \frac{1}{2} \frac{\p^2 \hat{\HH}_0}{\p \kappa \p z}\hat{A}=0.$$ As a result, we obtain a more simple equation 
\begin{equation} \label{parab_full}
\frac{\p u}{\p z} = \rmi k_0 \hat{H}[u], \qquad \hat{H} = -\bigg(k_0\frac{\p \hat{\HH}_0}{\p \kappa}\hat A\bigg)^{-1}\!\! \hat{\HH}_0 \hat A.
\end{equation}
This is our basic equation that describes the evolution of the scalar wave beam amplitude $u$ along $z$-axis,  including the effects of   diffraction, spatial dispersion and dissipation, as long as condition $|\p {u}/\p z| \ll \kappa |u|$ of the paraxial approximation holds. Except few very special cases, operator $\p \hat{\HH}_0/\p\kappa\approx\p \HH_{GO}/\p \kappa$ is  local, i.e. it may be approximated with multiplication by known function as long as the media is smooth compared to the wave length. In this case, the new amplitude is also defined by local transformation $u=\tilde u\sqrt{{k_0\p {\HH}_{GO}}/{\p \kappa}}$.

The simple form of Eq. \eqref{parab_full} suggests a clear physical interpretation of the evolution operator $\hat{H}$. Indeed, this is a longitudinal wave number expressed as a function of the transverse wave vector in an operator form:
\begin{equation}\label{UU}
k_0\hat{H} = k_z(\vec{r}, -\rmi \nabla_\perp  )-\kappa(z).
\end{equation}
Below we will show how to restore this operator from the WKB-solution $k_z(\vec{r}, \vec{k}_\perp )$ of an algebraic dispersion relation in a locally homogeneous medium.

%Note that the assumption of a small change of the group velocity over the beam cross-section can be broken by  strong inhomogeneity, such as an inclined incidence  of a wave beam onto a sharp vacuum-plasma boundary. In practice, such situations are usually accompanied by a pronounced reflection of radiation.

Our approach may be adopted to highly curved magnetic configurations, e.g. typical for radiation belts in Earth ionosphere or Solar flares, in a straightforward manner---just by introducing new coordinates with a curvilinear axis $z$ and taking into account the curvature when calculating a conjugated momenta operator. The similar approach was previously used for toroidal magnetic traps in which the axis $z$ was chosen along the reference geometric optics ray representing the center of the quasi-optical wave beam \cite{Balakin_2007a, Balakin_2008, BECCD}. However, using   geometric optics rays as a reference for the quasi-optical equation in open traps is not optimal because such rays may be strongly curved and even divergent inside the plasma column in most interesting cases \cite{X4,X9}.

\section{Evolution operator}

The main difficulty in the practical use of the quasi-optical equation \eqref{parab_full} is the uncertainty of the dielectric permittivity operator $\hat{\ee}$, which is necessary for definition of the evolution operator $\hat{H}$. Obviously, the operator $\hat{\ee}$  for any inhomogeneous linear media can be generated by some kernel function $\ee_{ij}(\vec{r},\vec{k})$, and such kernel can be introduced in infinite different ways. The most optimal (for our needs) way is proposed in quite rigorous manner in Ref.~\onlinecite{operator}, where the dielectric operator is represented as 
\begin{multline}\label{eq:ee_def}
\hat{\ee}_{ij} = 
% \frac12\sum_{n} \frac{1}{n!}\left \{\frac{\p^{n} \ee_{ij}}{\p \vec{p}\!^n}, \hat{\vec{p}}^n\right\},
\ee_{ij}+\frac{1}{2}\left (\frac{\p \ee_{ij}}{\p {k_\alpha}} \Delta\hat{{k}}_\alpha+\Delta\hat{{k}}_\alpha \frac{\p \ee_{ij}}{\p {k_\alpha}}\right)+\\ +\frac12\frac{1}{2!}\left (\frac{\p^{2} \ee_{ij}}{\p {k_\alpha} \p k_\beta} \Delta\hat{k}_\alpha\Delta\hat{k}_\beta+\Delta\hat{k}_\alpha\Delta\hat{k}_\beta \frac{\p^{2} \ee_{ij}}{\p {k_\alpha}\p {k_\beta}}\right)+\ldots=\\
=\frac12\sum_{n=0}^{\infty} \frac{1}{n!}\left (\frac{\p^{n} \ee_{ij}}{\p \vec{k}^n} \Delta\hat{\vec{k}}^n+\Delta\hat{\vec{k}}^n \frac{\p^{n} \ee_{ij}}{\p \vec{k}^n}\right).
\end{multline}
Here $\Delta\hat{\vec{k}}=\hat{\vec{k}}-\vec{k}=-\rmi\nabla-\vec{k}$.
For such operator, a Hermitian  kernel, $\ee_{ij}=\ee_{ji}^*$, always generates a Hermitian operator $\hat{\ee}$ in a sense of natural scalar product  $(a,b)=\int ab^*\: d\vec{\vec{r}}$. Correspondingly, an anti-Hermitian kernel, $\ee_{ij}=-\ee_{ji}^*$, generates an anti-Hermitian operator. This important non-trivial feature allows to preserve the energy conservation law when constructing the dielectric response of inhomogeneous media.  The kernel function  $\ee_{ij}$ is obtained as the sum of the amendments to the  dielectric permittivity tensor over the powers of characteristic inhomogeneity scale $l$: 
$$\ee_{ij} = \sum_{m=0}^{\infty} \ee_{ij}^{(m)}, \quad \ee^{(m)}_{ij} =\mathcal{O}((k_0 l)^{-m}).$$ 
As a reasonable approximation in a smoothly inhomogeneous media with $k_0 l \gg 1$,  one can only account the lowest term, $\ee_{ij} \approx \ee_{ij}^{(0)}$, which is indeed the dielectric permittivity tensor obtained for a ``locally homogeneous'' media, i.e. in the geometric optics approximation.

%Expressing in Eq.~\eqref{eq:ee_def} the differentiation with respect to spatial variables via a Fourier integral we obtain, after  some algebra, Eq.~\eqref{eq:e}. 
To find a Fourier representation, let us substitute 
$$E_j(\vec{r})=\int \rme^{\rmi \vec{k}' (\vec{r}-\vec{r}')} E_j(\vec{r}') \frac{d\vec{r}' d\vec{k}'}{(2\pi)^3},$$
$$E_j(\vec{r})\frac{\p^{n} \ee_{ij}(\vec{r},\vec{k})}{\p \vec{k}^n}=\int \rme^{\rmi \vec{k}' (\vec{r}-\vec{r}')} E_j(\vec{r}') \frac{\p^{n} \ee_{ij}(\vec{r}',\vec{k})}{\p \vec{k}^n}\frac{d\vec{r}' d\vec{k}'}{(2\pi)^3}.$$
into Eq.~\eqref{eq:ee_def}. Then 
\begin{multline*}%\label{eq:ee_def}
\hat{\ee}_{ij}[E_j(\vec{r}) ]= \int\frac12
\sum_{n=0}^{\infty} \frac{1}{n!}\frac{\p^{n}}{\p \vec{k}^n}\left ( \ee_{ij}(\vec{r},\vec{k})+ \ee_{ij}(\vec{r}',\vec{k})\right)\times\\  \times(\vec{k}'-\vec{k})^n\;\rme^{\rmi \vec{k}' (\vec{r}-\vec{r}')} E_j(\vec{r}') \frac{d\vec{r}' d\vec{k}'}{(2\pi)^3}.
\end{multline*}
Noting that summation over $n$ is just a Taylor series for $\ee_{ij}(\ldots,\vec{k}')$ we obtain Eq.~\eqref{eq:e}. It should be stressed that expressions \eqref{eq:e} and \eqref{eq:ee_def} remain valid for any kernel $\ee_{ij}(\vec{r},\vec{k})$, and for every linear inhomogeneous problem such kernel exists. The problem is, however, that except some trivial model cases, see e.g. Ref.~\onlinecite{operator}, this kernel is known only with the accuracy provided by geometric optics. 

The dielectric permittivity operator unambiguously defines the evolution operator in Eq.~\eqref{parab_full}. %  taking into account Eqs.~\eqref{eq_amplitude} and \eqref{UU}. 
However, we can cut the computations and obtain the same result by just noting that the evolution operator may be found   essentially in the same  way as the dielectric operator $\hat{\ee}$. So, the operator $\hat{H}$ is generated by the scalar function $H(\x,\q)$
\begin{equation}\label{eq:H_def}
\hat{H}[u] =\int \frac{H(\x,\q')+H(\x',\q')}{2} \rme^{\rmi \q' (\x-\x')} u(\x') \frac{d\x' d\q'}{(2\pi)^2} ,
\end{equation}
and as a practical approximation to $H(\x,\q)$ in a weakly inhomogeneous media, one can use a solution of the locally homogeneous problem. The only difference we should have in mind is that coordinate $z$ is a parameter here---there is no differentiation over $z$ in  operator $\hat{H}$, therefore instead of  6-dimensional phase-space $d\vec{r}'d\vec{k}'/(2\pi)^3$ we use 4-dimensional space $d\x'd\q'/(2\pi)^2$. Due to the term $\exp(\rmi \q' (\x-\x'))$, the evolutionary operator \eqref{eq:H_def} becomes local either in real space for an inhomogeneous medium without spatial dispersion with $H(\x)$, or in $\q$-space for a homogeneous dispersive medium  with $H(\q)$,  or for combination $H_1(\x)+H_2(\q)$. The latter  has been used in simulations for toroidal configurations \cite{Balakin_2008,BECCD}, 
but we find this approximation being not suitable for open traps. Thus, we must deal with a general form  $H(\x,\q)$ below. %, in this sense we call our present approach as ``non-perturbative''. 

As already mentioned, symmetric form of the integral representation  allows  to introduce  the dielectric response of an inhomogeneous medium that is  compatible with the energy conservation. In the present context, Eq.~\eqref{eq:H_def} always results in Hermitian operators for real kernel $H$ and anti-Hermitian operators for purely imaginary $H$. However, this feature is insufficient to provide a correct description of inhomogeneous dissipative medium with spatial dispersion. As shown in the next section, equation \eqref{eq:H_def} may give absurd results and, therefore, needs  modification. 

\section{Dissipative correction}

Let us decompose the linear operator $\hat{H}$ over Hermitian and anti-Hermitian parts
\begin{equation*}%\label{HH}%notused
\hat{H}=\hat{H}_H+\rmi\hat{H}_A,
\end{equation*}
where $\hat{H}_H\equiv(\hat{H}+\hat{H}^\dagger)/2$ and $\hat{H}_A\equiv (\hat{H}-\hat{H}^\dagger)/2\rmi$ are both Hermitian operators in the sense of the standard scalar product $\int ab^*\: d\vec{\x}$. 
It can be shown that with the same accuracy as  Eq.~\eqref{parab_full}, $\int |u|^2 d \vec{x}$ is proportional to the total wave energy flux in the cross-section $z$. Then, using the quasi-optical equation \eqref{parab_full} one can find the linear density of the absorbed power as
\begin{equation}\label{eq:P2}
\HP\equiv -\frac{\p }{\p z}\int |u|^2 d \vec{x}=2 k_0 \int u^*\hat{H}_A[u]\; d\vec{\x}.
\end{equation}
Hermitian operator in Eq.~\eqref{parab_full} conserves the energy flux, thus describing non-dissipative medium. As expected, the anti-Hermitian part of the evolution operator is responsible for the dissipation or pumping of the wave field. 

From the geometric optics limit, we know that energy dissipation corresponds  to a positive  sign of $\Imag H(\x,\q)$. Let us now consider the   eigenvalues $\lambda_A$ of  the ``dissipative'' operator $\hat{H}_A$  with the dissipative kernel $\Imag H(\x,\q)>0$ according to Eq.~\eqref{eq:H_def}. One may expect that all $\lambda_A>0$, however this is not true. Procedure \eqref{eq:H_def} guarantees only $\lambda_A^2>0$ since $\hat{H}_A$  is Hermitian, but not the positive sign of $\lambda_A$. In particular, we had negative $\lambda_A$  with a first try to apply Eq.~\eqref{eq:H_def} to model the resonant wave absorption in magnetized plasma. 

Fortunately, we have found a simple practical way to correct the evolution operator, which guarantees a definite sign of $\lambda_A$. To do this, we define $\hat{H}_A$ a square of some other Hermitian operator:
\begin{equation}\label{eq:Ha_G}
\hat{H}_A = \hat{G}\big[\hat{G}[u]\big],
\end{equation}
where operator $\hat{G}$ is constructed similar to Eq.~\eqref{eq:H_def} but with kernel $G(\x,\q)= \sqrt{\Imag H(\x,\q)}$%
\begin{equation*}\label{eq:G_def}
\hat{G}[u] = \int \frac{G(\x,\q')+G(\x',\q')}{2} \rme^{\rmi k_0 \q' (\x'-\x)} u(\x') \frac{d\x' d\q'}{(2\pi)^2}.
\end{equation*}
New operator $\hat{H}_A $ provides the same WKB limit as the old one. If the operator kernel is defined within the geometric optics approximation, then formal difference between both operators is of higher order than their accuracy. The modified quasi-optical equation \eqref{parab_full} then takes the following form
\begin{equation}\label{parab_full2}
\frac{\p u}{\p z} = \rmi k_0 \hat{H}_H[u] - k_0 \hat{G}[\hat{G}[u]].
\end{equation}

The discussed problem of the absorbed power  in an inhomogeneous medium with spatial dispersion may be illustrated with a simple one-dimensional example. Consider the following Hamiltonian kernel with positive imaginary part
\begin{equation*}%\label{samp:x2p2}%notused
H_A (x,k_x)= \alpha^2 x^2 k_x^2 \ge 0.
\end{equation*}
Corresponding non-modified evolution  operator \eqref{eq:H_def} is
$$  \hat{H}_A = - {\alpha^2} \left( x^2 \p_{xx} + 2x\p_x +1\right). $$
According to Eq.~\eqref{eq:P2}, the absorbed power density along the $z$ axis is
$$\HP= {2 \alpha^2}{k_0} \int \left(|x \p_x u|^2 - |u|^2 \right) dx.$$
The r.h.s. of this expression does not conserve the sign. For example, the Gaussian wave beam $u = u_0 \rme^{-(x-a)^2/w^2}$ corresponds to
$$\HP=\sqrt{\pi/ 8}\;{\alpha^2u_0^2k_0}(4a^2/w-w).$$
In particular, non-shifted beam with $a=0$, regardless of its width $w$, always results in a negative absorbed power. Shifted beam with $a=w/2$ propagates without dissipation. These strange solutions disappear for modified evolution  operator \eqref{eq:Ha_G}. In this case 
$$ \hat{G} = {\rmi \alpha}\left( x\p_x + \tfrac{1}{2} \right), \quad \hat{H}_A = - {\alpha^2} \left( x^2 \p_{xx} + 2x\p_x +\tfrac{1}{4} \right), $$
and the absorbed power density of the new operator possesses a positive sign:
\begin{multline*}
\HP= {2 \alpha^2}{k_0} \int \left(|x \p_x u|^2 - \tfrac{1}{4} |u|^2 \right) dx = \\ = {2 \alpha^2}{k_0} \int \left| x\p_x u + \tfrac{1}{2} u \right|^2 dx>0.
\end{multline*}
Note that the dissipation-free solution is still present, but in this case it corresponds to unbounded distribution $u \propto |x|^{-1/2}$ with an infinite power flux.

\section{Numerical solution }

The linear integro-differential equation \eqref{parab_full} involves time consuming integration in 4-dimensional space $d\x'd \q'$. Therefore, an algorithm economic  both in calculation of the r.h.s. and in number of $z$-steps is of major importance. We develop such algorithm having in mind essential  features of a beam propagation physics in resonant dispersive media.

As a starting point one can consider an explicit scheme in which a small but finite integration step  $\Delta$ over the evolution coordinate $z$ is defined as 
\begin{equation*} \label{sol:OE1}u(\x,z+\Delta) \approx  \hat S_H[u(\x,z)],
\end{equation*}
where
\begin{multline}\label{sol:OE}
 \hat S_H[u(\x,z)] \equiv  \int \exp \left( \rmi k_0 \Delta \frac{H(\x,\q')+H(\x',\q')}{2} \right) \times \\ \times \rme^{\rmi \q' (\x-\x')} u(\x',z) \frac{d\x' d\q'}{(2\pi)^2} .
\end{multline}
This is a generalization of the operator exponent technique studied in Refs.~\onlinecite{Balakin_2008, Fraiman95}.
One can find that this scheme results in an exact solution with any finite  $\Delta$  either for a homogeneous dispersive medium, $H(\q)$, or for an inhomogeneous medium without spatial dispersion, $H(\x)$. For a general case of a dispersive inhomogeneous medium, $H(\x,\q,z)$, this scheme converges to the exact solution for $\Delta \to 0$. 

The operator exponent method is rigorously justified only for dissipation-free media described by a Hermitian evolution  operator. In this case,  there is the effective numerical scheme for  calculating the operator \eqref{sol:OE} requiring no more than $\mathcal{O}(N^2)$ operations with $N$ being the number of mesh points in 2-dimensional space $\x$. In contrast to the traditional finite difference methods \cite{Samarsky},  there are no formal limitations on the step  $\Delta$  imposed by grid size in $\x$. Although the scheme is explicit, the integration  step is limited only by physical ``inhomogeneities'' of a medium, namely, by the value of  commutator between $\hat{H}(\x,\q)$ and $\hat{H}(\x',\q)$:
\begin{equation} \label{eq:applD}
\Delta \ll \frac{1}{k_0 \int |u|^2 d\x} \int \left| \frac{\p} {\p x_i}\hat{H}[x_i u] - x_i \frac{\p }{\p x_i}\hat{H}[u] \right|^2 d\x.
\end{equation}

However, operator \eqref{sol:OE} may result in a solution with an exponentially growing energy in inhomogeneous dissipative media. One of the reasons,  already described in the previous section, is that the dissipation in terms of geometric optics, $\Imag H(\x,\q)>0$, does not guarantee the positive-definiteness of the corresponding operator term $\hat{H}_A$. More generally, this is a known issue of finding a proper  approximation for physically correct wave absorption in media with a spatial dispersion\cite{Ginz, Brambilla, TokWest1, TokWest2, TokWest3, Balakin_diss}. In our case,  errors in the approximation of the Hermitian part (e.g. numerical errors due to violation of condition \eqref{eq:applD})  disturb only the wave beam phase, therefore, their influence is manifested over the large diffraction length. On the contrary,  errors in the approximation of the anti-Hermitian part directly affect the beam amplitude, resulting in the worst case in an exponential grows of an error. Our  experience tells that this problem is especially actual for microwave beams in the electron cyclotron frequency range, when the resonant dissipation varies dramatically both in coordinate and wave vectors spaces.

These problems are essentially solved with the modification  of anti-Hermitian part of the evolution operator suggested with Eq.~\eqref{eq:Ha_G}. For numerical solving of the corresponding quasi-optical equation \eqref{parab_full2}, we propose to use the ``split-step'' method, i.e. the sequential solutions for the separate Hermitian and anti-Hermitian parts. The integration step is then
\begin{equation*}\label{sol:split0}
u(\x,z+\Delta) \approx \exp({-k_0\Delta\hat{G}\hat{G}})\left[ {\hat{S}_{H}'[u(\x,z)]}\right]
\end{equation*}
with $\hat{S}_{H}'\approx \exp(\rmi k_0\hat{H}_H)$  is an approximation to the operator exponent of  the Hermitian part,  it is given by Eq.~\eqref{sol:OE} with $\hat{H}=\hat{H}_H$. 
Opposite to the Hermitian part, the  dissipation  operator does not fit the form suitable for an efficient computation of the operator exponent (just because it requires two iterations of $\hat{G}$). Calculation of the operator exponent  in this case requires about $\mathcal{O}(N^2 \max \lambda_G^2 k_0\Delta )$ number of steps, where $\lambda$ is an eigenvalue of matrix $\hat{G}$, which is typically too much for practical applications. 
Another way is to compute  $\exp(-k_0\Delta\hat{G}\hat{G})$ approximately by using conventional methods such as Adams or Runge--Kutta \cite{Samarsky}. Such calculations are also very time demanding, mainly because of large eigenvalues that may appear after a double iteration of $\hat{G}$ and the need to reduce the integration step $\Delta$ to provide a numerical stability. 

On the other hand, large positive eigenvalues correspond to quickly damped or even evanescent modes that typically do not represent physical interest. This makes it possible to replace the  operator exponent $\exp(-k_0\Delta\hat{G}\hat{G})$ with some approximate operator $\hat{D}$, which  correctly describes all weakly damped modes and  ensures exponential decay for heavily damped and non-propagating modes with large  eigenvalues.
To fulfill the first condition we require that both operators have the same asymptotic behavior at $\Delta \to 0$:
\begin{eqnarray*}
\exp(-k_0\Delta\hat{G}\hat{G})&=&\hat 1-k_0\Delta \hat{G}\hat{G} +O(\Delta^2),\\
\hat{D}&=&\hat 1-k_0\Delta \hat{G}\hat{G} +O(\Delta^2).
\end{eqnarray*}
To ensure that the second requirement, it is sufficient to ``cut'' the integral kernel of $\hat{D}$ at large $\Delta$. The simplest operator satisfying the above conditions is
\begin{equation}\label{sol:tanh}
\hat{D}[u] = u - \hat{T}[\hat{T}[u]],
\end{equation}
where operator $\hat{T}$ is constructed similar to Eq.~\eqref{eq:H_def} but with kernel $$T(\x,\q)=\phi( \sqrt{k_0 \Delta}\ G(\x,\q)),$$
$\phi(x)$ is an arbitrary one-dimensional function with the following properties: $\phi(x)\to x$ when $x\to 0$,  $\phi''(0)=0$ and $\phi(x)\to 1$ for $x\to \infty$. In our code we use the hyperbolic tangent $\phi=\tanh(x)$. Calculation of the operator \eqref{sol:tanh} reduces to a consistent application of the Fourier transform to the factorized kernel, so the computational complexity for $\hat{D}$ is only by a factor of two greater than that for $\hat{S}_{H}'$.

Now, we are ready to formulate the final explicit numerical scheme for quasi-optical equation \eqref{parab_full2}: one integration step is defined as
\begin{equation}\label{sol:split}
u(\x,z+\Delta) = \hat{D}[\hat{S}_{H}'[u]].
\end{equation}
In explicit form this reads
\begin{align*} %\label{sol:tanh}
&S(\x,\q)= \exp\left(\rmi {k_0 \Delta}\Real H(\x,\q,z)/2\right) , \\ 
&T(\x,\q)=\tanh \sqrt{k_0 \Delta \Imag H(\x,\q,z)}\:,\\
&u_1=   \int S(\x,\q') S(\x',\q')\:\rme^{ \rmi \q' (\x-\x')}\: u(\x',z) \frac{d\x' d\q'}{(2\pi)^2}\:, \\
&u_2=   \int \frac{T(\x,\q')+T(\x',\q') }{2} \:\rme^{\rmi \q' (\x-\x')}\: u_1(\x',z) \frac{d\x' d\q'}{(2\pi)^2}\:, \\
&u_3=   \int \frac{T(\x,\q')+T(\x',\q') }{2} \:\rme^{\rmi \q' (\x-\x')}\: u_2(\x',z) \frac{d\x' d\q'}{(2\pi)^2}\:, \\
&u(\x,z+\Delta)=u_1-u_3\:.
\end{align*}
Verification of this scheme on a set of model exactly solvable problems is published separately \cite{JETP}.

\section{Description of hot magnetized plasma in open magnetic trap}

In this section, we  define the kernel for quasi-optical evolution operator used in the numerical simulation of microwave plasma heating in open magnetic trap. Our model is based on the dispersion relation for electromagnetic waves propagating in a locally homogeneous hot magnetized plasma with Maxwellian distribution functions. This approach is justified for modern experiments in which the microwave energy is deposited into relatively dense plasma. Exactly in this case the electrodynamic quasi-optical effects become important, while the kinetic effects related to the rf-driven modification of the electron distribution function are not essential. For example, the Maxwellian distribution formed during ECR heating of bulk electrons  was proved at the GDT device by direct laser scattering measurements\cite{X8,X9,X10}. Kinetic effects are important in the opposite case of low density plasma, such as GAMMA-10 experiment\cite{V14} or ECR assisted start-up at the GDT\cite{Yack2016}, in which electrodynamics is usually well described by standard geometric optics or beam tracing, but dielectric response of non-Maxwellian electrons is calculated self-consistently by means of a quasi-linear Fokker--Planck code\cite{FP1}. 

In a vicinity of the fundamental electron cyclotron harmonic the dispersion relation of the Maxwellian plasma can be represented as follows \cite{tensor}:
\begin{multline} \label{eq:disp_EC}
n_\perp^2 \left[(\ee_+-\ee_\parallel)(\ee_--n^2) + (\ee_--\ee_\parallel) (\ee_+-n^2))\right] =\\= 2\ee_\parallel(\ee_+-n^2)(\ee_--n^2).
\end{multline}
where  $n_\parallel=c k_\parallel/\omega$ and $n_\perp=c n_\perp/\omega$ are parallel and perpendicular components of the index of refraction with respect to the external magnetic field, $n^2=n_\perp^2+n_\parallel^2$,
$$\ee_-=1+\frac{X Z(\zeta)}{n_\parallel \beta_e},\quad\ee_+ = 1-\frac{X}{1+Y}, \quad \ee_\parallel = 1-X,$$
$X=\omega_p^2/\omega^2$ and $Y=\omega_H/\omega$ are standard Stix parameters that determine the plasma density and the strength of the confining magnetic field through, respectively,  Langmuir and cyclotron frequencies of electrons, $\beta_e = \sqrt{2T_e/mc^2}$ is the normalized thermal  velocity  for electrons. All effects associated with the spatial dispersion and resonant dissipation at the fundamental harmonic are defined in the ``warm'' part in $\ee_-$ by so-called plasma dispersion function  \cite{Brambilla}
$$ Z(\zeta) = \exp({-\zeta^2} )\left( \rmi \sqrt{\pi} \sign \Real (n_\parallel) - 2 \int_0^\zeta \exp({t^2})\:dt \right) $$
with argument $\zeta =({1-Y})/({n_\parallel \beta_e})$. These formulas are obtained for a weakly relativistic plasma $\beta_e\ll 1$, a small Larmor radius $n_\perp\beta_e\ll Y$, and a quasi-longitudinal wave propagation $n_\parallel\gtrsim Y\beta_e $ that allows omitting the relativistic resonance broadening  compared to the non-relativistic Doppler shift. These conditions are definitely met, at least, in current experiments at the GDT device.

Having in mind applications to a long and axially symmetric devices, such as GDT, in the current implementation of the code we consider the external magnetic field directed  along the trap axis $z$ . This  simplifies the recovery of the quasi-optical Hamiltonian from the dispersion equation \eqref{eq:disp_EC}, however this assumption is not essential for our  technique and may be bypassed if necessary.

For strictly longitudinal magnetic field, the quasi-optical variables \eqref{eq:coords} are mapped to Stix variables as
$k_z= k_0 n_\parallel$, $\vec{q}=k_0 \vec{n}_\perp $. With this substitution, for each point in the real space and each transverse momentum $\vec{q}$ we determine a complex solution $n_\parallel(\x,\vec{q},z)$ of the local dispersion relation \eqref{eq:disp_EC}. Away from the EC resonance this solution is close to the roots of the biquadratic dispersion relation corresponding to a cold plasma limit, $\beta_e \to 0$. This allows us to choose the ``right'' root, matching the studied electromagnetic mode, and use it as a starting point for an iterative search for solution $n_\parallel$ of the transcendent equation \eqref{eq:disp_EC}. Additional control  for the right choice of the mode is provided by requirement of a smoothness of $n_\parallel$ over $\x$, in case of occasional sharp jumps  we use the values at neighboring grid cells to recover the correct mode.

Following Eq. \eqref{UU},  the complex kernel of the evolution operator $\hat H$ may be defined as 
\begin{equation} \label{eq:Hpl} H = n_\parallel(\x,\q,z)-\kappa(z)/k_0, \end{equation}
where $\kappa(z)=k_0\Real n_\parallel(\vec{\tilde x}, \vec{\tilde q}, z)$ is the carrier wave vector that corresponds  to  centers of mass for the wave field distribution  and its spectrum  in the cross-section $z$,
\begin{equation*}
\vec{\tilde x}=\frac{\int \x |u(\x,z)|^2 d\x}{\int  |u(\x,z)|^2 d\x}\;,\;\;\vec{\tilde q}=\frac{\int \q \left|\int u(\x,z) \rme^{\rmi\q\x} d\x\right|^2d\q}{(2\pi)^2\int  |u(\x,z)|^2 d\x}\;.
\end{equation*}
%Note that, since $\p H/\p n_\parallel\equiv 1$, the above definition guarantees a constant  group velocity  over a beam aperture.

Another problem specific for magnetic plasma confinement,  is evaluation how the absorbed microwave power is  distributed over a trap volume. Here one should keep in mind that this  power  rapidly redistributes over the magnetic flux surfaces. In this case, the most relevant characteristic of the power deposition is a one-dimensional distribution over the magnetic surfaces, usually referred to as a power deposition profile. This distribution may be found taking into account Eq.~\eqref{eq:P2}, as
\begin{equation}\label{pi_full}
P(\rho) = \frac{2 k_0}{l(\rho)} \int \Real (u^*\hat{H}_A[u]) \delta(\rho - \rho(\x,z)) \, d\vec{\x} \, dz,
\end{equation}
where the integral is taken over the whole volume occupied by a wave beam, equation $\rho(\vec{x},z)=\rho$ defines the magnetic surface,   $\rho$ without arguments is  the label of a magnetic surface,   
$$ l(\rho) = \int \delta(\rho-\rho(\vec{x},z_0)) \, d\vec{\x} $$ 
is the effective perimeter of a magnetic surface at some fixed cross-section $z=z_0$, for an axially symmetric magnetic configuration $l=2\pi\rho$.

The code \textsl{QOOT} (Quasi-Optics for Open Traps) is developed to solve the quasi-optical equation \eqref{parab_full2} with iterative procedure \eqref{sol:split} for the  evolutionary operator defined by kernel \eqref{eq:Hpl}. The code calculates the power deposition profile \eqref{pi_full} in the geometry of an open magnetic trap and visualizes the results. We use the Cartesian coordinates for the wave amplitudes and output of the results, and the internal cylindrical coordinates for the Hamiltonian taking advantage of the axial symmetry of the plasma configuration.
As mentioned in the introduction, the development is based on our previous quasi-optical code \textsl{LAQO}, created for toroidal magnetic traps \cite{Balakin_2008}. However, since conditions for the wave propagation in open mirror traps and toroidal traps are significantly  different, we have eventually developed an entirely new code.

%\section{Simulation of ECR plasma heating at the GDT device }

\begin{figure*}
\centering \includegraphics[width=0.9 \textwidth]{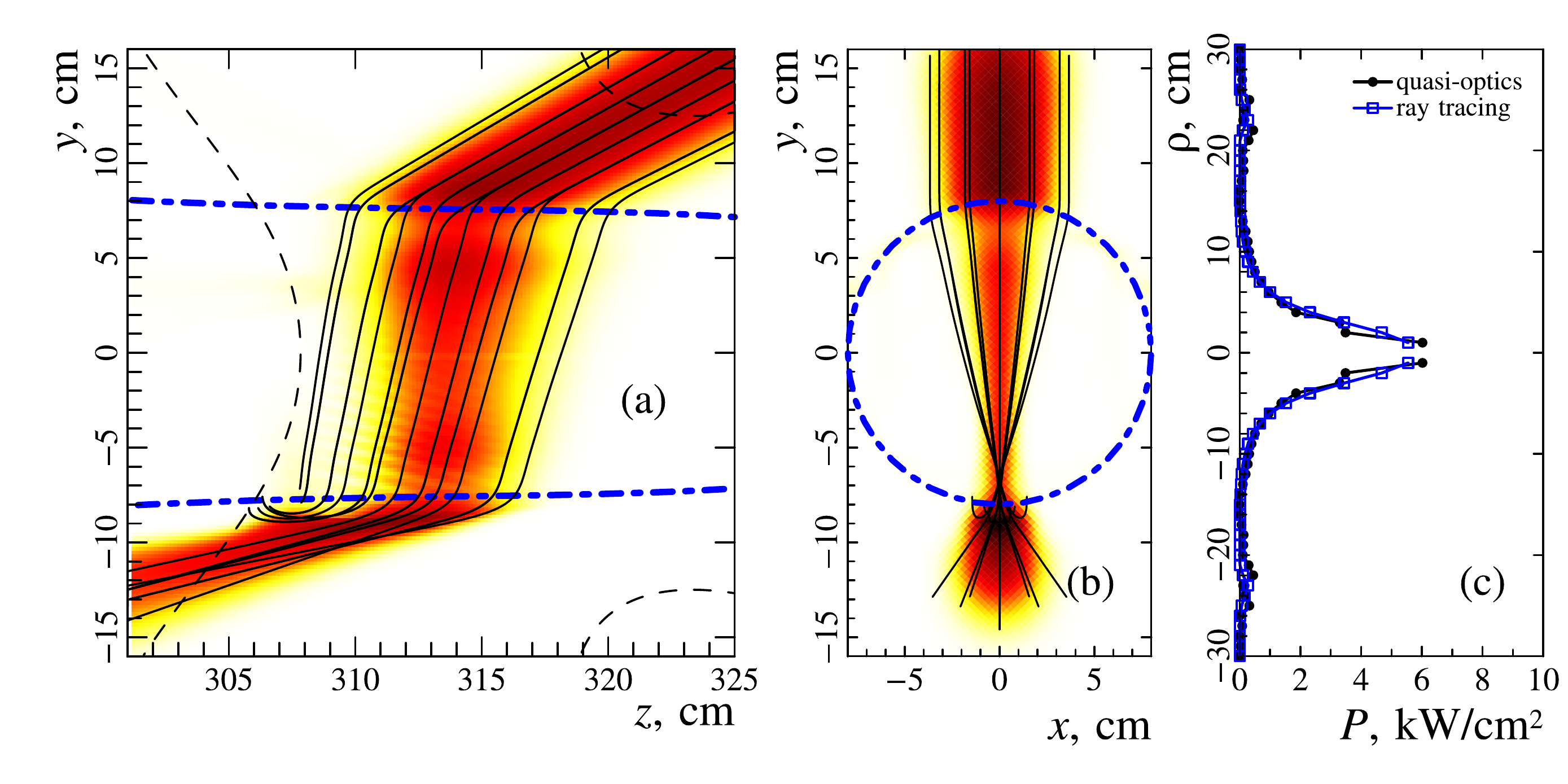}
\caption{(color online) Simulations of the ``narrow power deposition'' regime before the ECR heating. Distributions of the wave beam intensity  (a -- side view, b -- face view) and the power deposition profile related to the trap center (c). The solid lines in panels (a) and (b)  show the geometric optics rays, dashed and dotted curves indicate the EC resonance and the plasma boundary, respectively.  The efficiency of absorption is almost 100~\%.} \label{fig_qo1a}
\end{figure*}
\begin{figure*}
\centering \includegraphics[width=0.9 \textwidth]{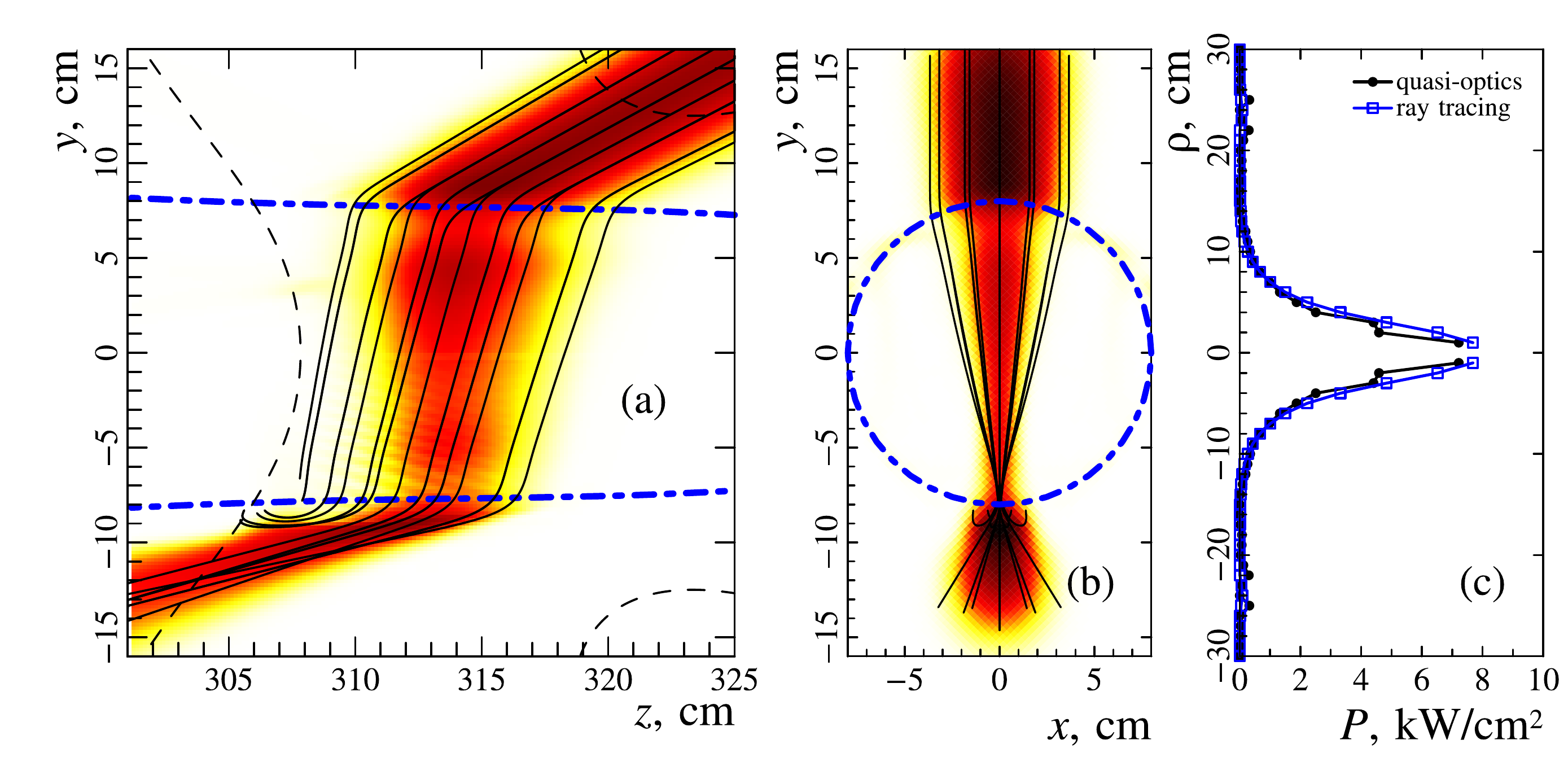}
\caption{(color online) Simulations of  the ``narrow power deposition'' regime after the ECR heating. 
The efficiency of absorption is almost 100~\%.} \label{fig_qo1b}
\end{figure*}
\begin{figure*}
\centering \includegraphics[width=0.9 \textwidth]{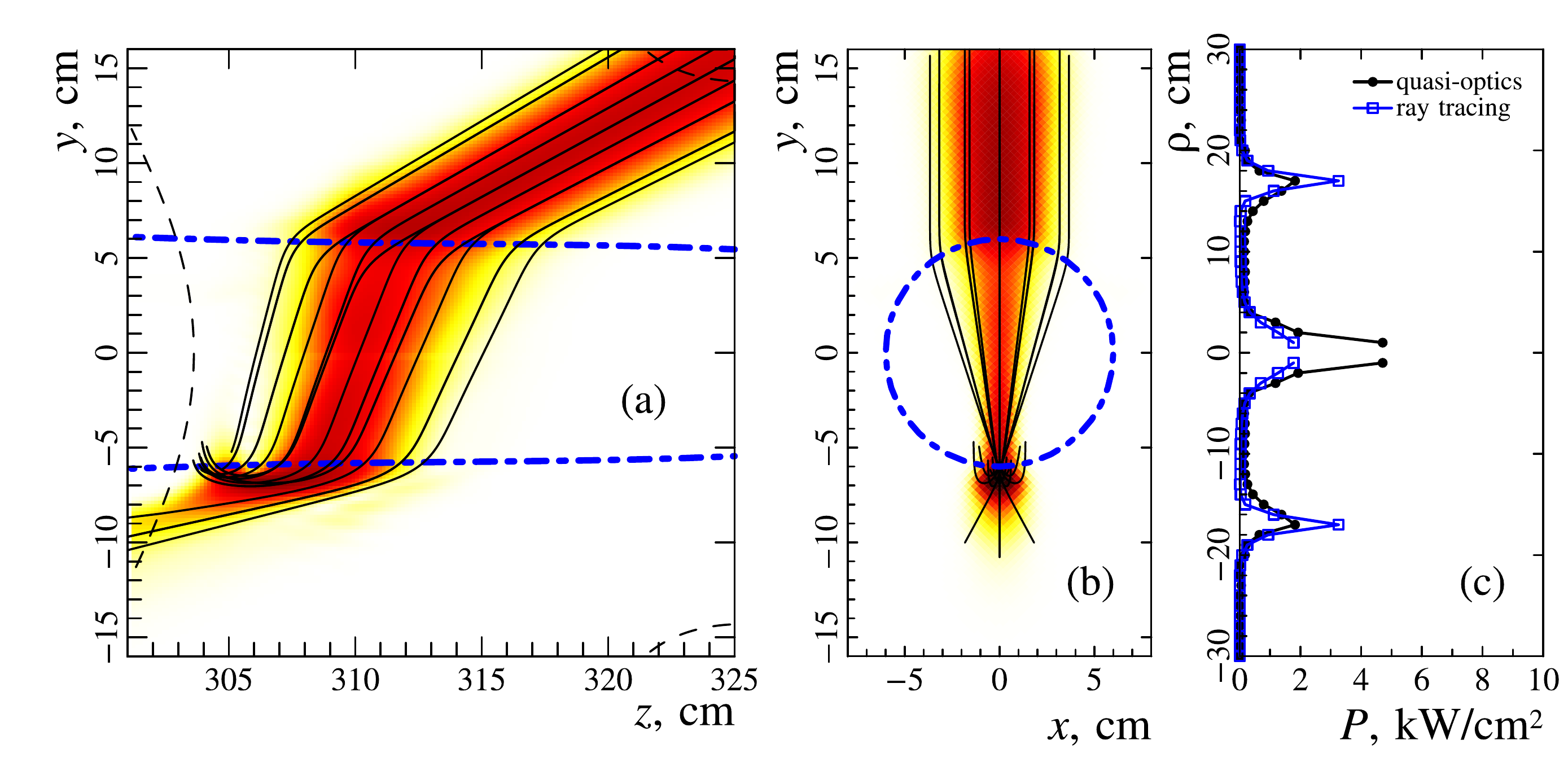}
\caption{(color online) Simulations of  the ``broad power deposition'' regime before the ECR heating. 
The efficiency of absorption is about 80~\%.} \label{fig_qo2a}
\end{figure*}
\begin{figure*}
\centering \includegraphics[width=0.9 \textwidth]{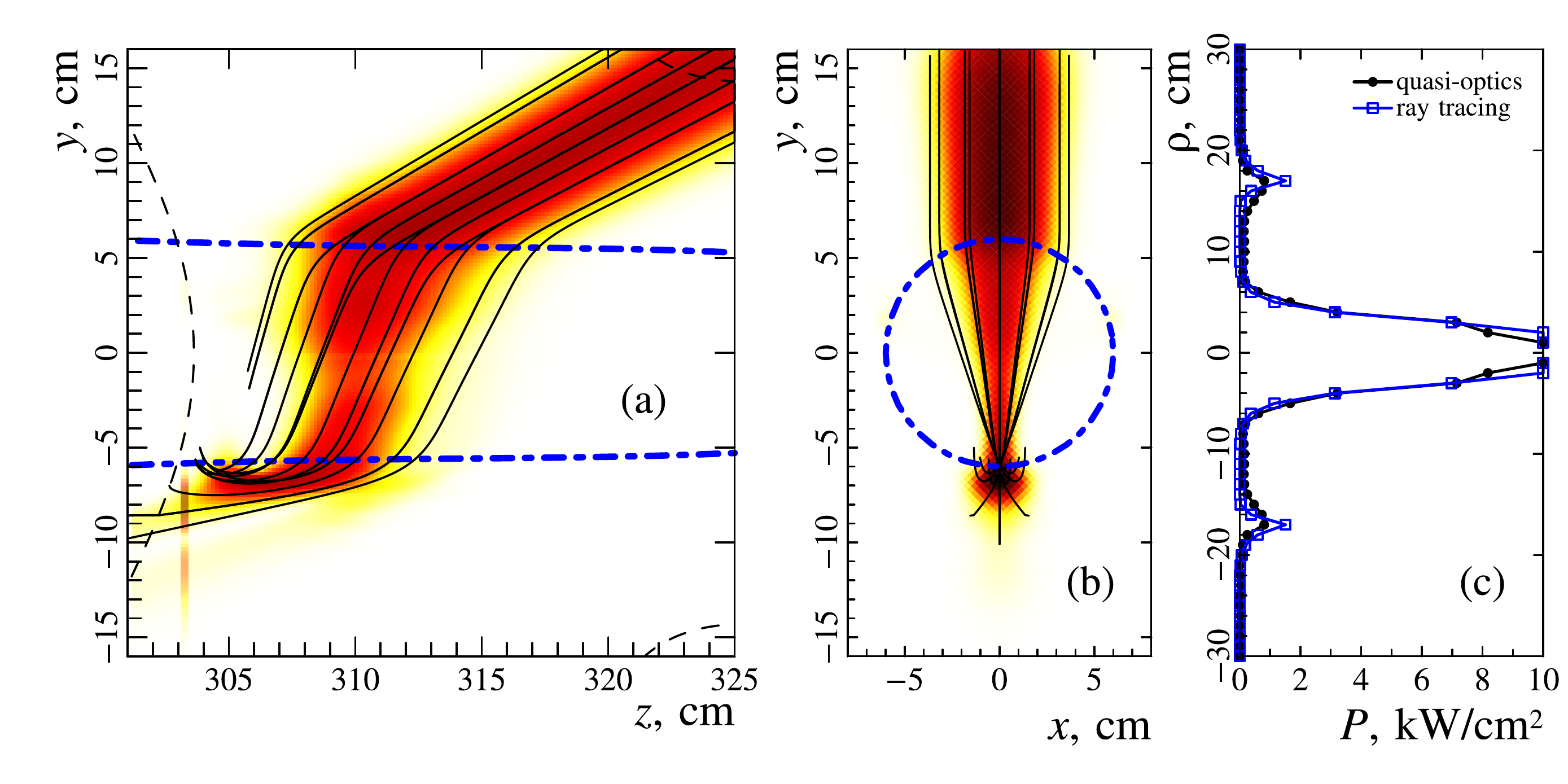}
\caption{(color online) Simulations of  the ``broad power deposition'' regime after the ECR heating. 
The efficiency of absorption is about 80~\%.} \label{fig_qo2b}
\end{figure*}
\begin{figure*}
\centering \includegraphics[width=0.9 \textwidth]{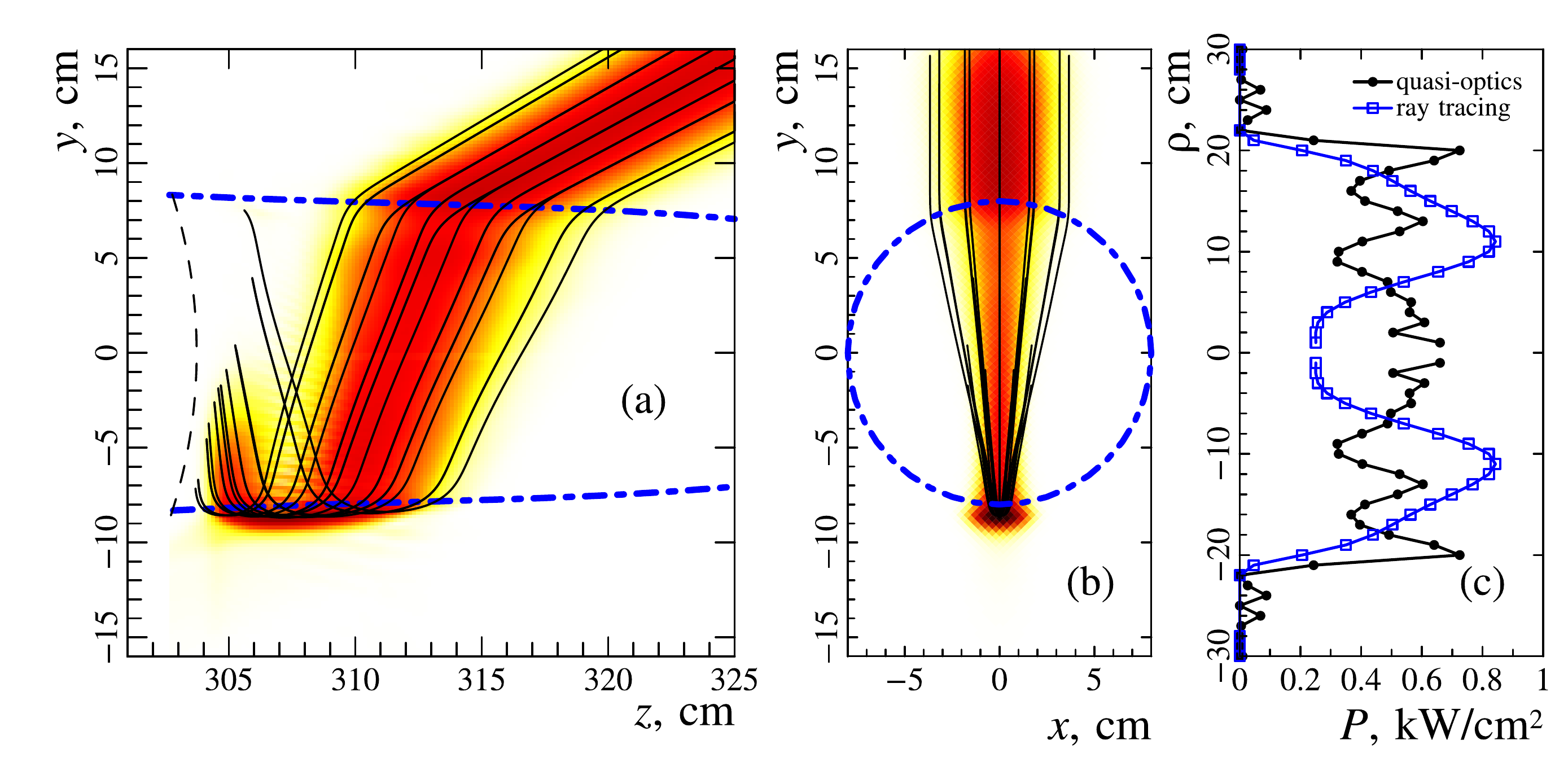}
\caption{(color online) Simulations of  the ``improved broad power deposition'' regime before the ECR heating. 
The efficiency of absorption is almost 100~\%.} \label{fig_qo3}
\end{figure*}

\section{Simulation of ECR plasma heating at the GDT device }

The basic ECR  heating scheme  at the GDT  relies on radiation trapping by a non-uniform plasma column \cite{X4}. This effect is caused by a dependence of the plasma refraction of the magnetic field strength. The radiation is launched through a side of the plasma column at a high magnetic field (close to the magnetic mirror). As a microwave beam propagates in plasma in the direction of the trap center, the magnitude of the magnetic field decreases resulting in  conditions for an internal reflection from the plasma-vacuum boundary. The plasma column acts as kind of waveguide, heterogeneous in both transverse and longitudinal directions, whereby the radiation is delivered to the ECR surface. The geometric optics may not be valid in the vicinity of the wave reflection  surface due to caustic formation, and near ECR  surface due to sharp inhomogeneous damping of the wave field. This casts doubt on the applicability of geometrical optics approximation. 

Therefore, the first task for the quasi-optical code is to check our previous results obtained with ray-tracing.  In particular, we repeated simulations aimed at optimization of the ECR heating efficiency at the GDT conditions. Previously, it has been found that the efficient ECR heating can be achieved in two distinguished regimes characterized by very different distributions of the absorbed microwave power \cite{X9, X10}. Switching between these regimes, as well as a fine-tuning of  the microwave power deposition  profile can be controlled, at  specific conditions of the GDT experiment,  by a relatively small adjustment of the external magnetic field in the close proximity of the EC resonance.

The results are shown in Figures \ref{fig_qo1a}--\ref{fig_qo3}. For  two-dimensional visualization of a quasi-optical beam in $(x,y)$ plane (face view along the trap axis) and $(y,z)$ plane (side view) we use the wave intensity integrated along the $z$ and $x$ axis, correspondingly:
$$J_z(x,y)=\int \chi\; |u|^2 dz,\quad J_x(y,z)=\int \chi\; |u|^2 dx.$$
Here $|u|^2$ characterizes the density of the energy flux along the $z$ axis, and $\chi=|\vec{v}_{\mathrm{gr}}|/({\vec{v}_{\mathrm{gr}}\vec{z}_0})$ projects that flux to the direction of the the group velocity $\vec{v}_{\mathrm{gr}}$. In explicit form
%\begin{multline*}%\left|\frac{\p\Real H(\vec{r}, \vec{p}) }{\p\vec{p}}\right|_{\vec{p}=(\vec{\tilde q},\kappa)}=\\=
$$\chi^2={ 1+\left(\frac{\p \Real k_z}{\p k_x}\right)^2_{\q=\vec{\tilde q}}+\left(\frac{\p \Real k_z}{\p k_y}\right)^2_{\q=\vec{\tilde q}} }. $$
%\end{multline*}
To improve the contrast, the logarithmic color scale is used corresponding to the levels  of $\ln J$, which allowed us to visualize the wave field in caustics. The power deposition profiles $P(\rho)$ are calculated using Eq.~\eqref{pi_full}, applied at the central cross-section where the confining magnetic field has its minimum.

Figures \ref{fig_qo1a} and  \ref{fig_qo1b} show the results of simulation, the wave beams and the power deposition profiles, in the  ``narrow power deposition'' regime. In this regime the record values of the electron temperature (up to 1 keV) have been achieved in the GDT experiments. First and second figures correspond, respectively, to states before and after the ECR heating modeled for the experimentally measured  plasma density and electron temperature profiles. For comparison, we show the ray-tracing results. In this case, the power deposition profiles obtained by two different methods coincide quite well. Some discrepancy is observed on the periphery of the plasma column. This corresponds to the absorption of radiation after reflection in the vicinity of the caustic surface where geometric optics is violated (caustics are indicated as crossing of neighboring ray trajectories). Note, that the geometric optics rays  reproduce  the quasi-optical beam satisfactorily both before and inside the caustic region.

Figures \ref{fig_qo2a} and  \ref{fig_qo2b} show the same plots for the  ``broad power deposition profile'' regime. In the experiments in this mode, we observe  a pronounced  increase of a total plasma energy related to an improved confinement of the hot ions. There is a better agreement between the quasi-optical and geometro-optical  power deposition profiles since the ratio of microwave power deposited in the vicinity of the caustic is much lower. As well as the ray-tracing, the quasi-optical simulations predict incomplete absorption of the microwave power in this regime.

Figure \ref{fig_qo3} shows predictions for the new regime with  ``improved broad power deposition''. This mode has not been yet  demonstrated  experimentally; its implementation will be eventually possible after up-grade of the mirror magnetic coil of the GDT system which is currently on progress. All plots correspond to the stage before the microwave heating because the experimental data on plasma profiles after the heating is not available. In this mode, an extended region of the caustic is formed, and the ratio of the power deposition after the caustic is much higher than in the regimes discussed previously. Therefore, the power deposition profiles predicted by the quasi-optics and the ray-tracing  vary considerably. We are aware that the quasi-optical modeling  provides a more adequate description, but confirmation of this statement requires further studies and experiments, the results of which will be published separately.

Finally, it can be concluded that geometric optics seems to be a reasonable agreement with the more accurate quasi-optical simulations for the most heating scenarios that are already realized in the GDT experiment. However, this conclusion does not exclude a possible impact of spatial dispersion on the resonance dissipation and of diffraction losses near the caustic surfaces in future experiments with more optimized heating scenarios.

\section{Summary}

The quasi-optical model of propagation of wave beams in high-temperature magnetically confined plasmas, developed earlier for toroidal  traps, is generalized over open magnetic systems. The specifics of microwave heating in modern open traps require substantial improvement of the early quasi-optical theory, associated with more accurate description of the effects of spatial dispersion in the region of resonant wave dissipation. As a result, a new form for the quasi-optical equation is proposed, see Eq.~\eqref{parab_full2}. Basing on this equation the universal code \textsl{QOOT} is developed for simulation of electron cyclotron plasma heating, which allows to resolve the diffraction, dispersion and aberration effects in the propagation and absorption of electromagnetic wave beams in open traps.

The  code is used to verify the results of optimizing the efficiency of the ECR heating in the large mirror trap GDT, previously obtained by using ray-tracing  within geometric optics approximation. First quasi-optical simulations  justify the possibility to control the radial distribution of the deposited microwave power  by local modification of the magnetic field in near the EC resonance  and, in particular, the ability of effective on-axis heating the electron component at the GDT conditions. The influence of wave caustics on a localization of the deposited microwave power is demonstrated.

\begin{acknowledgments}
This work was supported by the Russian Science Foundation (grant No~14-12-01007). 
The authors would like to thank Dr. Alexander Solomakhin from the Budker Institute for his support with magnetic configurations and ray-tracing modeling for the GDT.
\end{acknowledgments}

\end{document}